\documentclass[%
reprint,
superscriptaddress,
nofootinbib,
 amsmath,amssymb,
 aps,
 prd,
]{revtex4-2}

\usepackage{graphicx}
\usepackage{dcolumn}
\usepackage{bm}
\usepackage{xcolor}
\usepackage{subfigure}
\usepackage{tabularx}

\begin{document}

\title{Measurements of low-energy nuclear recoil quenching factors for Na and I recoils in the NaI(Tl) scintillator}

\author{S.~H.~Lee}
\affiliation{University of Science and Technology~(UST), Daejeon 34113, South Korea}
\affiliation{Center for Underground Physics, Institute for Basic Science~(IBS), Daejeon 34126, South Korea}
\author{H.~W.~Joo}
\affiliation{Department of Physics and Astronomy, Seoul National University, Seoul 08826, South Korea}
\author{H.~J.~Kim}
\affiliation{Department of Physics, Kyungpook National University, Daegu 41566, Republic of Korea}
\author{K.~W.~Kim}
\email{kwkim@ibs.re.kr}
\affiliation{Center for Underground Physics, Institute for Basic Science~(IBS), Daejeon 34126, South Korea}
\author{S.~K.~Kim}
\affiliation{Department of Physics and Astronomy, Seoul National University, Seoul 08826, South Korea}
\author{Y.~D.~Kim}
\affiliation{Center for Underground Physics, Institute for Basic Science~(IBS), Daejeon 34126, South Korea}
\affiliation{University of Science and Technology~(UST), Daejeon 34113, South Korea}
\author{Y.~J.~Ko}
\affiliation{Center for Underground Physics, Institute for Basic Science~(IBS), Daejeon 34126, South Korea}
\author{H.~S.~Lee}
\email{hyunsulee@ibs.re.kr}
\affiliation{Center for Underground Physics, Institute for Basic Science~(IBS), Daejeon 34126, South Korea}
\affiliation{University of Science and Technology~(UST), Daejeon 34113, South Korea}
\author{J.~Y.~Lee}
\affiliation{Department of Physics, Kyungpook National University, Daegu 41566, Republic of Korea}
\author{H.~S.~Park}
\affiliation{Korea Research Institute of Standards and Science, Daejeon 34113,South Korea}
\author{Y.~S.~Yoon}
\affiliation{Korea Research Institute of Standards and Science, Daejeon 34113,South Korea}

\date{\today}

\begin{abstract}

Elastic scattering off nuclei in target detectors, involving interactions with dark matter and coherent elastic neutrino nuclear recoil (CE$\nu$NS), results in the deposition of low energy within the nuclei, dissipating rapidly through a combination of heat and ionization.
The primary energy loss mechanism for nuclear recoil is heat, leading to consistently smaller measurable scintillation signals compared to electron recoils of the same energy. The nuclear recoil quenching factor (QF), representing the ratio of scintillation light yield produced by nuclear recoil to that of electron recoil at the same energy, is a critical parameter for understanding dark matter and neutrino interactions with nuclei.
The low energy QF of NaI(Tl) crystals, commonly employed in dark matter searches and CE$\nu$NS measurements, is of substantial importance. Previous low energy QF measurements were constrained by contamination from photomultiplier tube (PMT)-induced noise, resulting in an observed light yield of approximately 15 photoelectrons per keVee (kilo-electron-volt electron-equivalent energy) and nuclear recoil energy above 5\,keVnr (kilo-electron-volt nuclear recoil energy). Through enhanced crystal encapsulation, an increased light yield of around 26 photoelectrons per keVee is achieved.
This improvement enables the measurement of the nuclear recoil QF for sodium nuclei at an energy of 3.8 $\pm$ 0.6\,keVnr with a QF of 11.2 $\pm$ 1.7\,\%. Furthermore, a reevaluation of previously reported QF results is conducted, incorporating enhancements in low energy events based on waveform simulation. The outcomes are generally consistent with various recent QF measurements for sodium and iodine.

\end{abstract}

\maketitle

\section{\label{sec:intro}Introduction}

Currently, weakly interacting massive particles (WIMPs), deemed as one of the most promising dark matter candidates~\cite{lee77,jungman96}, have been extensively explored by numerous experiments, yet with no conclusive findings~\cite{LUX-ZEPLIN:2022xrq,XENON:2017vdw}. This has prompted a shift in focus towards the low-mass dark matter realm, driving increased interest in low energy events~\cite{Kimura:2023fzj,SuperCDMS:2017nns,COSINE-100:2021poy,XENON:2019zpr}. This growing interest is relevant not only to dark matter studies but also to neutrino experiments, especially those involving coherent elastic neutrino-nucleus scattering (CE$\nu$NS)~\cite{Abdullah:2022zue}. In these experiments, neutrinos interact with target materials, depositing less than a few keV nuclear recoil energies (keVnr), emphasizing the importance of understanding low energy nuclear recoils.

Scintillating detectors detect nuclear recoil by measuring electron recoil energy, a process that involves distinct scintillation phenomena different from those associated with electron recoil energy~\cite{knoll}. Consequently, the quenching factor (QF), indicating the ratio of nuclear recoils to electron recoils, plays a crucial role in searches for WIMPs and CE$\nu$NS. Variations in the QF have substantial implications for the interpretation of physics outcomes~\cite{Ko:2019enb,NEON:2022hbk}.

The NaI(Tl) crystal is widely used in applications such as WIMP dark matter searches~\cite{Bernabei:2020mon,Amare:2021yyu,Angloher:2016ooq,Antonello:2020xhj,Fushimi:2015sew} and neutrino experiments~\cite{NEON:2022hbk}. Positive signals reported by the DAMA/LIBRA experiment~\cite{Bernabei:2010,Bernabei:2018yyw,Bernabei:2020mon} have attracted attention, but interpretations of these signals remain uncertain~\cite{Savage:2008er,Ko:2019enb,ParticleDataGroup:2022pth}. Although the COSINE-100 data have ruled out specific interpretations~\cite{COSINE-100:2021xqn}, model-independent data analysis is inconclusive~\cite{COSINE-100:2021zqh}. The ANAIS-112 experiment presents inconsistent results with DAMA/LIBRA at about 3$\sigma$ level~\cite{Amare:2021yyu}, but a potential difference in QFs is suggested by DAMA/LIBRA~\cite{Bernabei:2020mon}.

DAMA/LIBRA reported QF measurements using neutrons from a $^{252}$Cf source~\cite{BERNABEI1996757}. The measured responses were compared with the simulated neutron energy spectrum to obtain QF values of 0.3 for sodium and 0.09 for iodine, assuming independence from the energy of the recoiling nuclei.
Recently, various QF measurements from refined methods of NaI(Tl) crystals with mono-energetic neutron beams have been reported~\cite{Collar:2013gu,sabre1,Awe:2018fei,Joo:2018hom,Bignell:2021bjx,Cintas:2024pdu}. In these measurements, the detection of an elastically scattered neutron at a fixed angle relative to the incoming neutron beam direction provides energy-dependent QF values significantly lower than DAMA/LIBRA's QF results, approximately 0.1 at 10\,keVnr for sodium.
Although different methods of nuclear recoil calibration may contribute to this discrepancy, the crystal dependence of the QF values is not definitively excluded. Since DAMA/LIBRA reported their annual modulation results with a 0.75\,keVee energy threshold~\cite{Bernabei:2021} corresponding to 2.5\,keVnr (assuming 0.3 sodium QF of DAMA/LIBRA~\cite{BERNABEI1996757}), understanding low energy nuclear recoil QF is essential.

The calibration of low energy nuclear recoil in NaI(Tl) crystals is highly necessary, yet no measurements below 5\,keVnr currently exist, primarily due to noise induced by photomultiplier tubes (PMTs) contaminating a few photoelectron events corresponding to about 0.5\,keVee energy. We have developed a novel crystal encapsulation technique aimed at enhancing the light collection efficiency of NaI(Tl) crystals by directly attaching PMTs to the crystal, resulting in nearly 50\,\% increased light yield~\cite{NEON:2022hbk}. This technique has been applied to the NEON experiment, measuring CE$\nu$NS at a reactor, yielding a consistently high light yield above 22 photoelectrons per keVee (NPE/keVee)~\cite{NEON:2022hbk}. For a small-sized crystal with dimensions of 1.8\,cm $\times$ 1.8\,cm $\times$ 1.4\,cm, a comparable crystal encapsulation method was adopted, achieving a 26\,NPE/keVee high light yield. This allowed us to measure low energy events below 0.5\,keVee.

This paper presents measurements of the nuclear recoil QFs of a NaI(Tl) crystal utilizing a high light yield NaI(Tl) crystal detector. Simulating the scintillation waveform of NaI(Tl) crystals~\cite{Ko:2022pmu,Choi:2024ziz}, the characteristics of low energy scintillation events are effectively investigated while mitigating PMT-induced noise events. The event selection efficiency and the trigger efficiency are assessed through simulated signal samples, reaching a 0.2\,keVee analysis threshold. This threshold enables us to measure the low energy QF as low as the nuclear recoil energy of 3.8\,keVnr for sodium and 6.1\,keVnr for iodine.

It is noteworthy that our prior measurements, conducted with a NaI(Tl) crystal having a light yield of approximately 15\,NPE/keVee~\cite{Joo:2018hom}, demonstrated similar energy-dependent tendencies. However, the sodium QF values at 10\,keVnr were nearly 30\% smaller compared to measurements by other research groups~\cite{sabre1,Cintas:2024pdu}. With an enhanced understanding of low energy scintillation events and a precise calibration of incident neutron energy, we identified a bias toward lower QF values in our earlier measurement. Consequently, a thorough reexamination of the data analysis for the previously reported data was performed. The new measurement, deploying a high light yield NaI(Tl) detector, along with the reevaluated results of the previous measurements, yields QF values consistent with those reported by other research groups for both sodium~\cite{sabre1,Cintas:2024pdu,Bignell:2021bjx} and iodine~\cite{Collar:2013gu}.

\section{\label{sec:neutronE} Time of flight (TOF) measurements for incident neutron energy calibration  }
Mono-energetic neutrons are produced through the deuteron-deuteron nuclear fusion reaction using a DD109 neutron generator (Adelphi Technology, Inc.)~\cite{DD109}. The generator tube was successively shielded by borated polyethylene (40\,cm thickness) and high-density polyethylene (40\,cm thickness), which complies with safety regulations. Neutrons were emitted through a 3.5-cm-diameter hole in the shield at a 90$^{\circ}$ angle relative to the deuteron beam. The entire experimental setup was aligned in the direction of the neutron beam.

The incident neutron energy was determined using a $^{3}$He detector~\cite{Manolopoulou_2012,THOMAS20101178} positioned in front of the collimated hole when the DD109 generator was initially installed, resulting in a measured neutron beam energy of 2.43 $\pm$ 0.03\,MeV at an 85 keV deuteron energy. 
However, for previous QF measurements, a 60\,keV deuteron energy was utilized to address higher trigger rates caused by higher neutron and x-ray fluxes at higher deuteron energy. Given the reported near independence of neutron energy from the deuteron energy at the 90$^{\circ}$ angle~\cite{csikai1987handbook,10.1063/1.3586154}, the assumption was made that the neutron beam energy remains at 2.43 $\pm$ 0.03\,MeV even when the deuteron energy is reduced to 60\,keV.
Nevertheless, due to the volume of the titanium target inside the generator and the NaI(Tl) crystal~\cite{DD109}, there could be a certain angle distribution slightly different from 90$^{\circ}$. For a systematic understanding of the incident neutron energy, we employed the neutron energy calibration method based on time-of-flight (TOF) measurement.

\begin{figure}[htb]
    \begin{center}
        \includegraphics[width=0.8\columnwidth,keepaspectratio]{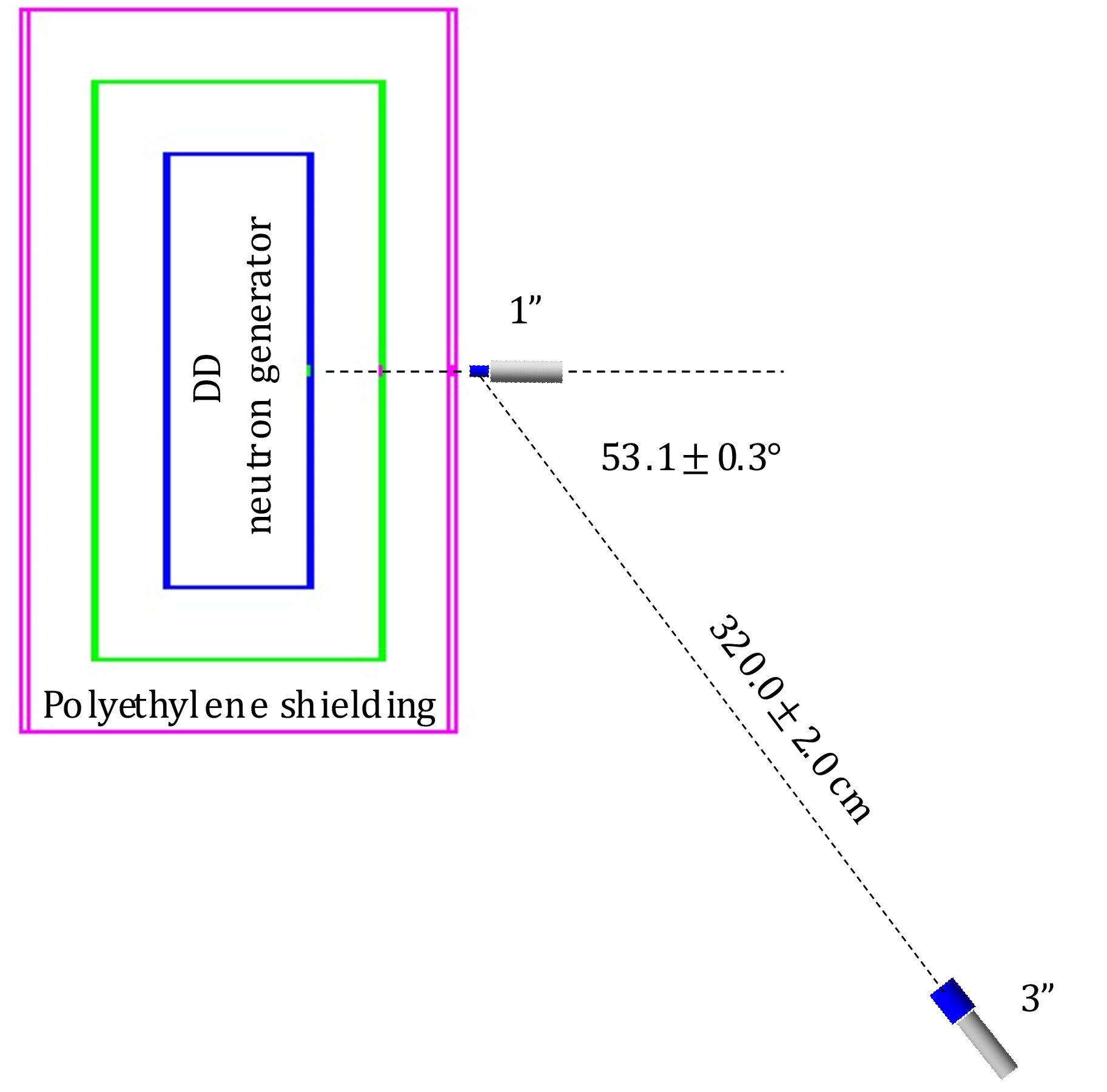}
    \end{center}
		\caption{Experimental setup of the incident neutron energy measurements using the TOF method. Neutrons are coming out from the deuteron-based generator inside the polyethylene shielding. One-inch diameter LS detector (LS1) is placed in front of the neutron beam direction and a three-inch diameter LS detector (LS2) is placed far from the one-inch detector to measure the TOF of the scattered neutrons.
        }
    \label{fig:tofsetup}
\end{figure}

Two liquid scintillator (LS) detectors were installed using EJ-301 from Eljen Technology, previously used for neutron tagging in QF measurements~\cite{Joo:2018hom}. Adjacent to the collimated hole of the neutron generator, a one-inch diameter LS detector, named LS1, was situated. Additionally, a second LS detector (LS2) with a three-inch diameter was placed at  a 53.1 $\pm$ 0.3$^{\circ}$ relative to the neutron beam direction and 320.0 $\pm$ 2.0\,cm distance, as depicted in Fig.~\ref{fig:tofsetup}.

If a neutron interacts with LS1 and subsequently with LS2, a TOF of the scattered neutron can determine the scattered neutron energy $E_{n'}$. 
\begin{equation}
    E_{n'} = \frac{M_{n}}{2}\left(\frac{d}{c\Delta t}\right),
\label{eq:nprime}
\end{equation}
where $\Delta t$ is the TOF between two LS detectors, $d$ is a distance between two LS detectors, and $c$ is the speed of light. 
The incident neutron energy $E_n$ is the sum of the recoil energy $E_{rec}$ in the LS1 detector and $E_{n'}$, 
\begin{equation}
E_{n}=E_{rec}+E_{n'},
\label{eq:En}
\end{equation}
where
$E_{rec}$ is determined as follows:
\begin{equation}
    E_{rec} = \frac{2(1+A-\cos^{2} \theta - \cos\theta \sqrt{A^{2}-1+\cos^{2}\theta})}{(1+A)^{2}}E_{n}.
\label{eq:Erec}
\end{equation}
Here, $\it{A}$ is the atomic mass number of the recoiled nuclei, and $\theta$ is the angle between the neutron beam direction and the scattered neutron direction.

\begin{figure}[htb]
    \begin{center}
    \includegraphics[width=1.0\columnwidth,keepaspectratio]{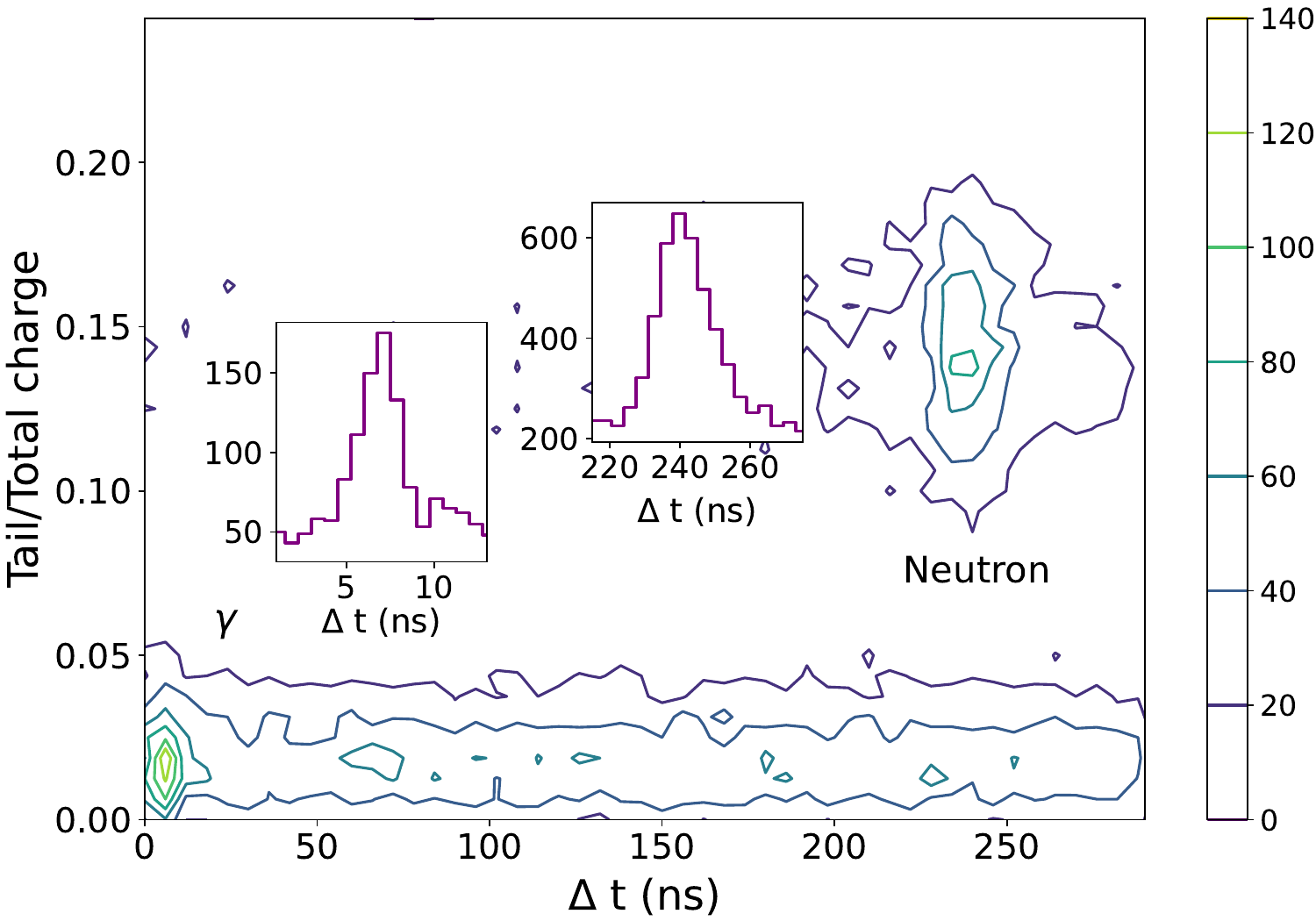}
    \end{center}
		\caption{
  The TOF ($\Delta t$) of the scattered neutron in two LS detectors as a function of the PSD parameter, the tail-to-total charge ratio. Neutron and $\gamma$ events are well separated with the PSD parameter as well as TOF. One-dimensional distribution of TOF ($\Delta t$) for each particle is placed in the insets. 
  }
    \label{fig:lse100}
\end{figure}

The TOF with a deuteron energy of 100\,keV was first measured. To assess the TOF of the neutron and $\gamma$ events, we differentiate between the induced particles based on their pulse shapes, taking advantage of the longer decay time characteristic of the neutron events~\cite{COSINE-100:2018jke}. The pulse shape discrimination (PSD) parameter is defined as a ratio of tail charge to total charge~\cite{Joo:2018hom,COSINE-100:2018jke}. As illustrated in Fig.~\ref{fig:lse100}, coincident neutron events and $\gamma$ events are effectively identified based on the PSD parameter and different TOFs.
The measured TOF values are found to be 240.8 $\pm$ 0.3\,ns and 6.8 $\pm$ 0.1\,ns for neutron and $\gamma$ events, respectively. With a distance between the two LS detectors of 320.0 $\pm$ 2.0\,cm, the TOF for $\gamma$ events should be 10.7 $\pm$ 0.1\,ns. Therefore, the measured TOF is corrected by adding 3.9 $\pm$ 0.1\,ns. The corrected neutron TOF is then determined to be 244.7 $\pm$ 0.3\,ns.
Using Eq.~\ref{eq:En} with hydrogen interaction in the LS detector, the incident neutron energy is evaluated as 2.48 $\pm$ 0.07\,MeV, where the uncertainty accounts for the uncertainties in the distance and angle.

\begin{figure}[htb]
    \begin{center}
        \includegraphics[width=1.0\columnwidth,keepaspectratio]{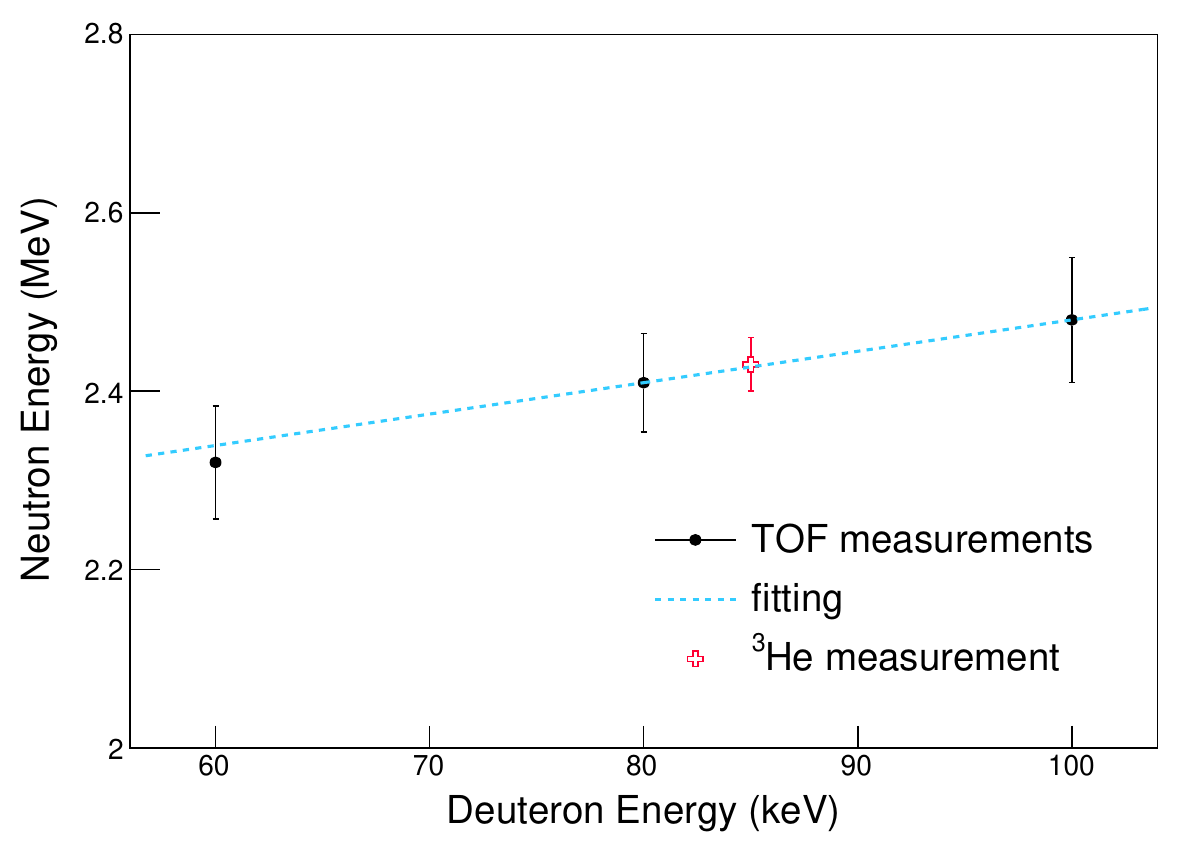}
    \end{center}
		\caption{The measured neutron energies from the TOF measurements as a function of deuteron energy (black dots) are shown together with $^{3}$He detector measurement (red cross) from Ref.~\cite{Joo:2018hom}. The empirically chosen linear fit to the data is represented by a blue dashed line.}
    \label{fig:nede}
\end{figure}

The incident neutron energy was measured based on the TOFs of two LS detectors for two more deuteron energies, specifically at 60\,keV and 80\,keV. The resulting measured neutron energies are 2.32 $\pm$ 0.06\,MeV and 2.41 $\pm$ 0.06\,MeV for 60\,keV and 80\,keV, respectively. All results are presented in Fig.~\ref{fig:nede}.
Notably, the incident neutron energies exhibit a dependence on the deuteron energies. 

\section{\label{sec:new}QF measurement with an improved crystal encapsulation}

\subsection{\label{sec:setup}Experimental setup}
\begin{figure}[htb]
\centering
\subfigure[]
{
    \includegraphics[width=0.6\columnwidth,keepaspectratio]{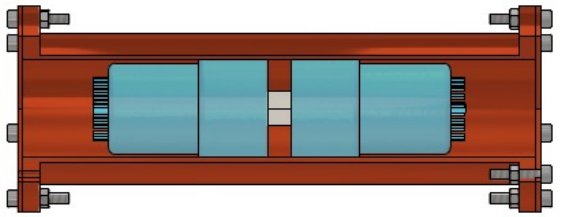}
    \label{fig:encap}
}
\centering
\\
\subfigure[]
{
    \includegraphics[width=1.0\columnwidth,keepaspectratio]{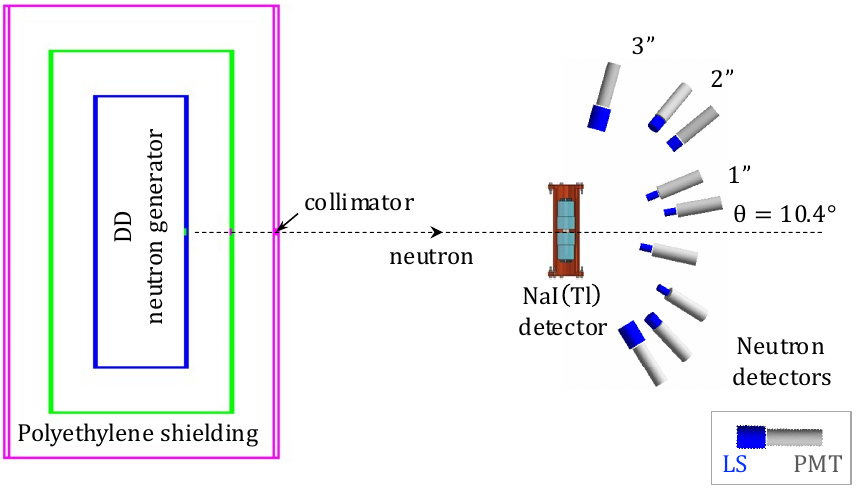}
    \label{fig:setup}
}
\caption{(a) Schematic design of NaI(Tl) detector encapsulation, where the PMTs are directly attached to the crystal. (b) Experimental setup of the QF measurement. Monoenergetic neutrons from the DD109 neutron generator are directed towards the NaI(Tl) detector. Various sizes of nine LS detectors are strategically positioned around the NaI(Tl) crystal for neutron tagging purposes. The one-inch LS detector at 10.4$^{\circ}$ was specifically installed for the low energy nuclear recoil measurement below 5\,keVnr.}
\label{fig:nai}
\end{figure}

To maximize light collection of a NaI(Tl) crystal, a small-sized crystal with dimensions of 1.8\,cm $\times$ 1.8\,cm $\times$ 1.4\,cm is directly attached to two 3-in. R12669 PMTs (Hamamatsu Photonics), as illustrated in Fig.~\ref{fig:encap}. The crystal is grown from the WIMPScint-III powder, produced by Alpha Spectra Inc., and originates from the same ingot as the COSINE-100  crystal 6 and 7~\cite{Adhikari:2017esn}. Instead of having a quartz window to encapsulate the crystal with copper or aluminum case, only a 1\,mm thin optical pad EJ-560 (Eljen Technology) connects the crystal to the PMTs optically. This technique has already been proved to enhance light collection~\cite{Choi:2020qcj,NEON:2022hbk} efficiency. 
The crystal-PMTs assembly is housed in an aluminum casing with a thickness of 5\,mm.

The encapsulated crystal is oriented towards the neutron beams in front of the collimated hole, positioned 1.8\,m away as depicted in Fig.~\ref{fig:setup}. We positioned the detector to expose its smallest face, measuring 1.8\,cm $\times$ 1.4\,cm, in the neutron beam direction to minimize multiple scattering of the incident neutrons.
For this new measurement, a fixed deuteron energy of 100\,keV was used, maintaining stable operation conditions of the DD109 generator. This corresponds to an incident neutron energy of 2.48 $\pm$ 0.07\,MeV as shown in Fig.~\ref{fig:nede}. 
A 5\,cm thick lead shield was added outside the polyethylene shield of the DD109 generator to mitigate radiated x-ray or other generator-related backgrounds. Additionally, for background reduction, a 10\,cm lead layer and a 20\,cm polyethylene layer were employed to shield the NaI(Tl) crystal in the direction of the neutron beam. A 3\,mm opening was incorporated for controlled exposure.

Nine EJ-301 LS detectors were deployed surrounding the NaI(Tl) detector for neutron tagging.
These LS detectors come in three different sizes: four with a 1-in. diameter, three with a 2-in. diameter, and two with a 3-in. diameter, arranged from the beam direction to outer angles. Specifically, to measure low energy nuclear recoils below 5\,keVnr for sodium, one of the 1-in. diameter detectors is placed in the colinear beam direction at 10.4$^{\circ}$. The arrangement of the DD109 generator, the NaI(Tl) crystals, and the nine neutron tagging detectors is illustrated in Fig.~\ref{fig:setup}.

A trigger is initiated when a signal from the PMTs in the NaI(Tl) crystal, corresponding to one or more photoelectrons, occurs in each PMT within a 200\,ns time window. In addition, hits from at least one of the neutron tagging detectors are required for triggering. The triggered signals are then recorded as 8\,$\mu$s waveforms with a 500\,MHz sampling rate, utilizing the same data acquisition system employed in the COSINE-100 experiment~\cite{COSINE-100:2018rxe}.



\subsection{\label{sec:anal}Analysis}

\subsubsection{\label{sec:nai_ly}NaI(Tl) crystal calibration and light yield}
The electron-equivalent energy scale of the NaI(Tl) crystal is linearly calibrated using a single $\gamma$ line of 59.54\,keV, achieved through exposure to a $^{241}$Am radioactive source.
To estimate the light yield, we first identify a cluster of photoelectrons~\cite{Kims:2005dol}, and then assess the light output using the ratio between the charge of the 59.54\,keV events and that of a single photoelectron~\cite{Adhikari:2017esn}.
In this study, a high light yield of 26.0 $\pm$ 0.7\,NPE/keVee was obtained, facilitated by the novel crystal encapsulation technique~\cite{Choi:2020qcj}.
This represents around a 70\% increase compared to the average values of 15\,NPE/keVee obtained from COSINE-100 detectors with the previous design~\cite{Adhikari:2017esn}, enabling exploration below 0.5\,keVee energy.

We routinely conducted $^{241}$Am calibration measurements every day to monitor the stability of light yield and correct gain variations. Throughout the approximately one-month measurement period, the detector module consistently maintained a stable and high light yield.

\subsubsection{\label{sec:nai_data}Nuclear recoil event selection}
\begin{figure}[htb]
    \begin{center}
        \includegraphics[width=1.0\columnwidth,keepaspectratio]{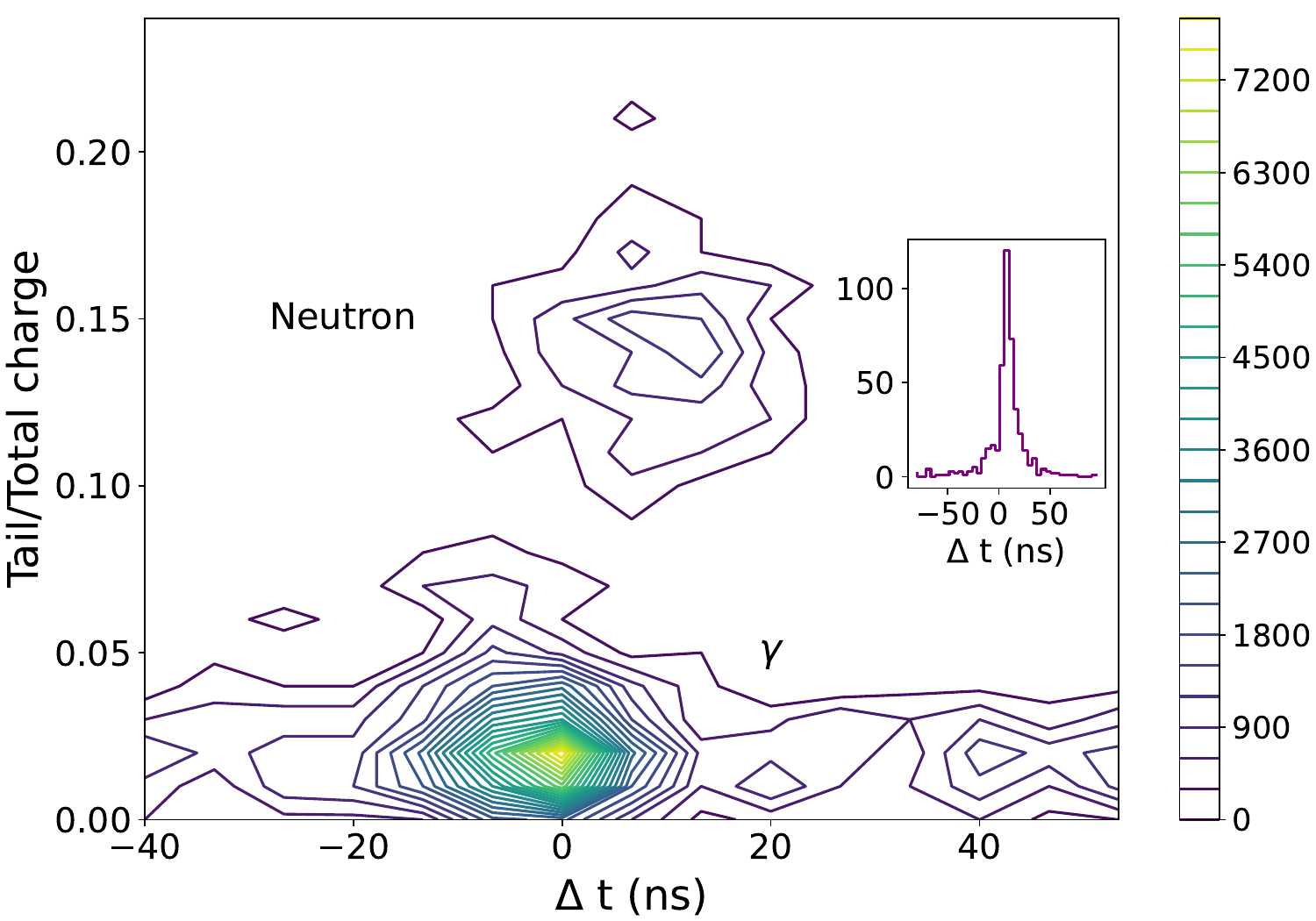}
    \end{center}
		\caption{The tail to total charge ratio of the LS neutron tagging detectors plotted against the TOF ($\Delta\it{t}$) between the NaI(Tl) and the LS neutron detectors. Nuclear recoil events, characterized by a longer decay time in the LS detector, exhibit higher values of the tail to total charge ratio. The inset illustrates the one-dimensional TOF for the neutron event, highlighting the clear coincident timing between NaI(Tl) and neutron tagging detectors. 
  }
    \label{fig:nPSD}
\end{figure}

The discrimination of nuclear recoil events from electron/$\gamma$ background was achieved by utilizing nine neutron tagging detectors. Two parameters were introduced to identify nuclear recoil events: the charge ratio of the 50 to 200\,ns (tail) to the 0 to 200\,ns (total), a robust discrimination parameter~\cite{COSINE-100:2018jke} leveraging the longer decay time of nuclear recoil events, as depicted in Fig.\ref{fig:nPSD}. Additionally, neutron events scattered in the NaI(Tl) crystal and tagged by the LS neutron tagging detectors showed a clear coincident timing in TOF between the NaI(Tl) crystal and the neutron tagging detector, as shown in Fig.\ref{fig:nPSD}. The combination of these two parameters facilitates the straightforward selection of isolated neutron candidate events.

\subsubsection{\label{sec:nai_noise} Waveform simulation}
In understanding the behavior of signals in low energy regions, we implemented a waveform simulation developed to characterize NaI(Tl) scintillation events recorded with the COSINE-100 and NEON DAQ systems~\cite{COSINE-100:2018rxe} as detailed in Ref.~\cite{Choi:2024ziz}. The simulation generates scintillation signals of the NaI(Tl) crystal, tuned with measured data to account for characteristics ranging from single photoelectrons to the scintillation properties of the NaI(Tl) crystal, including rising and decaying times, as well as triggering and digitization aspects from the DAQ system.

The waveform simulation was fine-tuned using the selected nuclear recoil events discussed earlier. Extensive validation of the simulation was conducted, covering the shape of single photoelectrons to the reconstructed parameters of NaI(Tl) scintillation events, as presented in Ref.~\cite{Choi:2024ziz}. An example of the meantime parameter, shown in Fig.\ref{fig:evtsel}, demonstrates excellent agreement between scintillation events in the data and the simulation. Subsequently, the data from waveform simulation was utilized to evaluate signal efficiencies based on trigger and event selection criteria.

\subsubsection{\label{sec:nai_noise}PMT-induced noise rejection}
While coincident neutron scattering events are selected based on the aforementioned parameters, this sample may still contain PMT-induced noise events. These events arise from incident neutrons scattering in PMTs or crystal encapsulation material, and the scattered neutrons are subsequently tagged by the neutron tagging detectors. Neutron scattering in the PMTs or encapsulation materials can generate $\beta$/$\gamma$ particles that directly strike PMT glasses, producing Cherenkov light. In this scenario, PMT-induced noise events in the NaI(Tl) crystals and isolated neutron events in the LS neutron tagging detectors are recorded.

\begin{figure}[htb]
    \begin{center}
        \includegraphics[width=1.0\columnwidth,keepaspectratio]{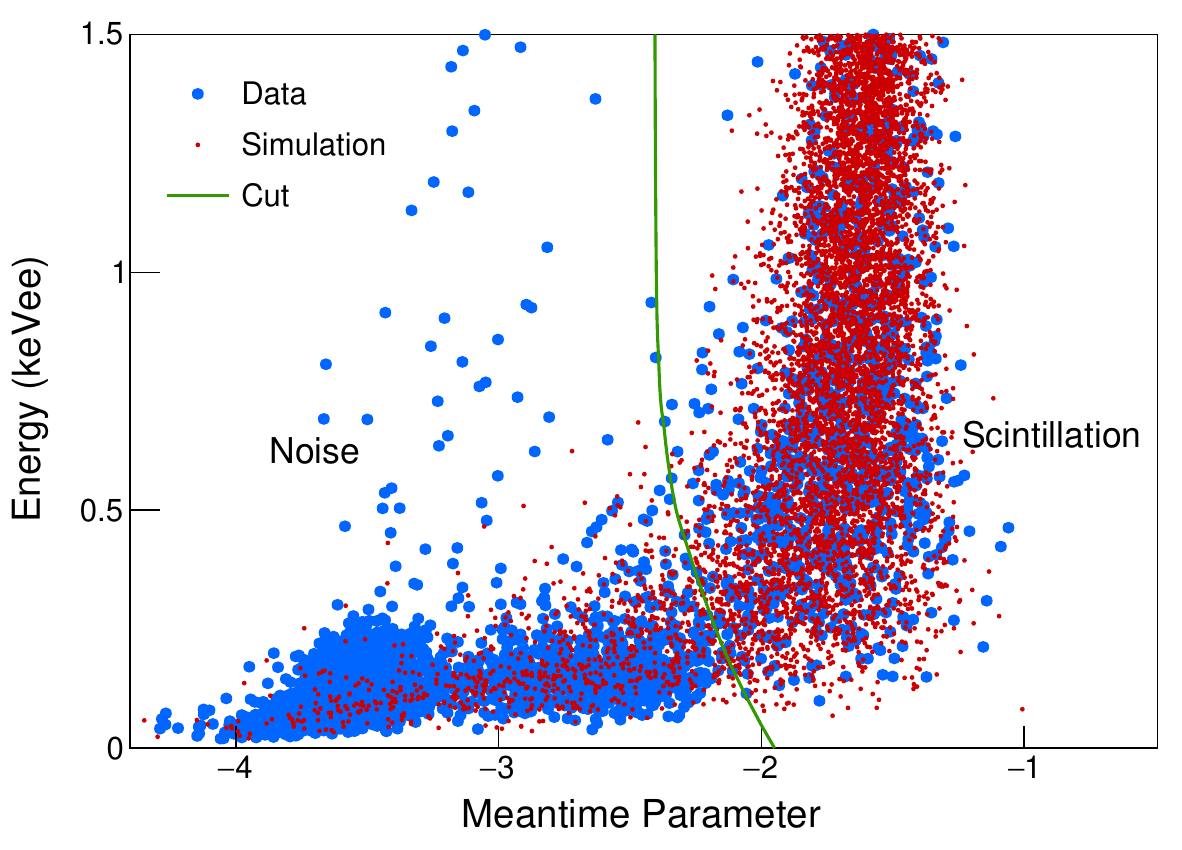}
    \end{center}
        \caption{
        Distributions of the meantime parameter for the measured data (blue points), which had a selection for the neutron scattering events, and the waveform simulation data (red points) are presented. The green-solid line indicates criteria for event selection to remove the PMT-induced noise events. 
        }
    \label{fig:evtsel}
\end{figure}

To discriminate PMT-induced noise from NaI(Tl) crystal's scintillation events, we introduced a meantime parameter developed for event selections in COSINE-100 data~\cite{COSINE-100:2020wrv}. The meantime parameter represents the logarithm of the combined charge-weighted mean decay time of each PMT, distinguishing events based on pulse shape, such as fast-decaying PMT-induced noise events from Cherenkov radiation in the PMT glasses. Figure~\ref{fig:evtsel} illustrates the meantime parameter as a function of energy for the neutron-tagged events.
PMT-induced noise events in this data are clearly distinguishable when compared with the waveform simulation data. It is noteworthy that contamination of PMT-induced noise is more critical at low energy regions. Using the selection criteria (solid line in Fig.~\ref{fig:evtsel}) for scintillation events, we can efficiently remove PMT-induced noise events for energies above 0.2\,keVee, corresponding to 5\,NPE.

\subsubsection{\label{sec:qf_calc} QF extraction}
\begin{figure}[htb]
    \begin{center}
        \includegraphics[width=1.0\columnwidth,keepaspectratio]{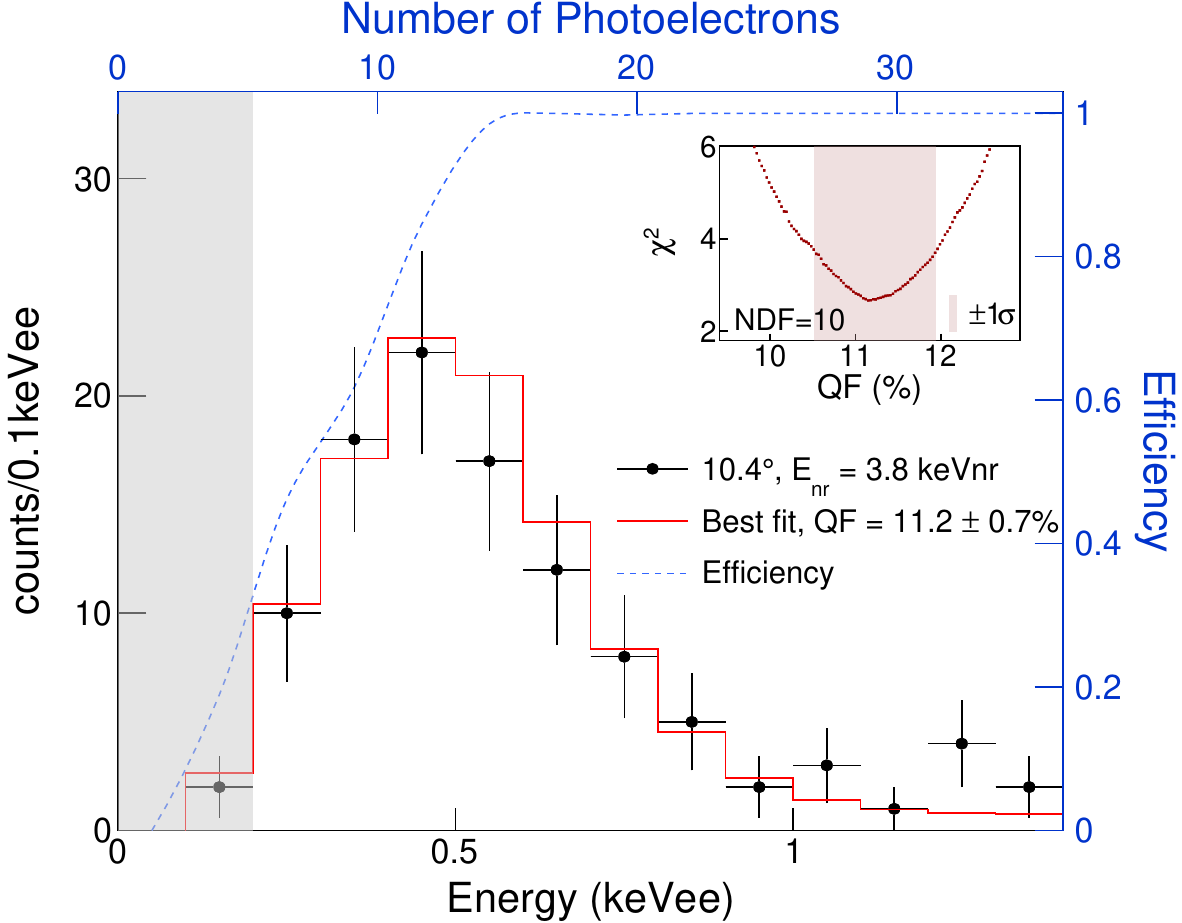}
    \end{center}
		\caption{The nuclear recoil energy spectrum of the NaI(Tl) crystal tagged by the neutron detector at 10.4$^\circ$ corresponding to 3.8 $\pm$ 0.6\,keVnr for sodium is shown with black-dot points. The best fit spectrum of QF (11.2\,\%) from the simulated spectra is overlaid with the red-solid line. 
  An inset shows the $\chi^2$ distribution as a function of the QF value with $\pm$1\,$\sigma$ range (orange band) representing the uncertainty. The total efficiency (blue-dashed line) was evaluated from the waveform simulation data and was applied to the simulated energy spectra.  }
    \label{fig:QF_temp}
\end{figure}

Events having coincidences with the neutron detectors, passing nuclear recoil selections, and clearing noise rejection cuts are applied to measure the QFs of the NaI(Tl) crystal.
The selected data for the fixed angle of the neutron tagging detector are compared with simulated energy spectra for various QF values. The {\footnotesize GEANT}4-based simulation includes mono-energetic neutron generation from the deuteron plate of the DD109 generator, the flight of neutrons to the NaI(Tl) crystal, scattering, and tagging of the scattered neutron with the neutron tagging detectors, with the geometry of the neutron measurement setup shown in Fig.~\ref{fig:nai}.
Efficiencies from the trigger and event selection are evaluated using waveform simulation data, as illustrated in Fig.\ref{fig:QF_temp}, and are applied to the simulated energy spectra. The evaluated efficiency is about 0.3 at 0.2\,keVee ($\approx$5\,NPE) and reaches unity above 15\,NPE, corresponding to about 0.6\,keVee.
The nuclear recoil energy from the collinear angle of 10.4$^{\circ}$ is evaluated as 3.8 $\pm$ 0.6\,keVnr. Minimization of $\chi^2$ between the measured energy spectrum and the simulated spectra from various QF values is performed. 
The best fit value is obtained from the minimum $\chi^2$, and the uncertainty is derived from the $\Delta \chi^2$\,=\,1 range (inset) as 11.2 $\pm$ 0.7\,\% as shown in Fig.~\ref{fig:QF_temp}.
The relative event rates between the data and the simulation at each scattering angle consistently agree, indicating no potential miscalculation of QF due to low energy event selection~\cite{Collar:2010ht}. 

We extract the QF values for both sodium and iodine from nine neutron tagging detectors. Due to relatively low energy nuclear recoils and low QF values, iodine QFs are evaluated for scattering angles above 33.0$^{\circ}$. Table~\ref{tab:QFresult} summarizes the measured QFs of sodium and iodine for nine neutron tagging detectors at different angles. Measured QFs strongly correlate with the nuclear recoil energy, with smaller values for lower recoil energies. The iodine QF is considerably smaller than for sodium recoils, similar to previous reports~\cite{Collar:2013gu,Joo:2018hom}.

\begin{table}[htb]
\caption{Summary of the QF measurement with the high light yield NaI(Tl) detector. Nuclear recoil energies ($E_{nr}$) at each scattering angle are taken from the {\footnotesize GEANT}4 simulation, accounting the deuteron plate structure, geometric structures of the NaI(Tl) crystals and the neutron tagging detectors. Uncertainties from neutron energies, efficiencies, and detector geometries are included.}
  \begin{center}
    \bgroup
    \def\arraystretch{1.3}
      \begin{tabular}{ccccc}
      \hline      \hline
	Scattering	&  \multicolumn{2}{c}{ Na } & \multicolumn{2}{c}{ I } \\
			      			\cline{2-5}
	angle ($^{\circ}$)	&           $E_{nr}$ (keVnr)           &           QF (\%)           &           $E_{nr}$ (keVnr)           &           QF (\%)           \\
      \hline
        10.4 &   3.8 $\pm$ 0.6 & 11.2 $\pm$ 1.7 & $-$ & $-$ \\
        12.7 &   5.6 $\pm$ 0.8 & 13.8 $\pm$ 1.6 & $-$ & $-$ \\
        24.6 &  19.8 $\pm$ 1.9 & 15.6 $\pm$ 1.1 & $-$ & $-$ \\
        33.0 &  33.6 $\pm$ 2.9 & 19.7 $\pm$ 1.2 & 6.1 $\pm$ 0.5 & 4.9 $^{+ \text{0.6}}_{- \text{1.8}}$ \\
        39.3 &  48.5 $\pm$ 3.7 & 18.7 $\pm$ 0.8 & 8.9 $\pm$ 0.7 & 5.0 $\pm$ 0.5 \\
        47.0 &  67.2 $\pm$ 6.1 & 19.7 $\pm$ 1.2 & 12.3 $\pm$ 1.1 & 4.7 $\pm$ 0.6 \\
        50.7 &  76.9 $\pm$ 6.9 & 20.1 $\pm$ 1.2 & 14.1 $\pm$ 1.3 & 5.3 $\pm$ 0.5 \\
        58.7 & 100.7 $\pm$ 9.2 & 22.7 $\pm$ 1.2 & 18.2 $\pm$ 1.7 & 5.9 $\pm$ 0.5 \\
        74.6 & 151.2 $\pm$ 14.2 & 24.5 $\pm$ 1.5 & 27.5 $\pm$ 2.8 & 6.8 $\pm$ 0.5 \\
       \hline
       \hline
    \end{tabular}
    \egroup
  \end{center}
    \label{tab:QFresult}
\end{table}
\section{\label{sec:reanal}Re-examine the previous QF measurement}
The decision to reexamine the previous QF measurement, reported in Ref.~\cite{Joo:2018hom}, was prompted by the identification of an incident neutron energy calibration issue, as detailed in Sec.~\ref{sec:neutronE}. During this reexamination, we employ a methodology similar to that used for the new measurement, incorporating advanced event selection and efficiency evaluation based on waveform simulation.

The incident neutron energy from the previous measurement has been re-evaluated and adjusted to 2.34 $\pm$ 0.06\,MeV, differing from the initially reported value of 2.43 $\pm$ 0.03\,MeV, as depicted in Fig.~\ref{fig:nede}. This revision leads to a lower neutron energy and consequently, a smaller nuclear recoil energy in the NaI(Tl) crystals. For instance, at the lowest scattering angle of the previous measurement (13.2$^{\circ}$), the originally reported nuclear recoil energy of 5.8 $\pm$ 0.4\,keVnr has been corrected to 5.5 $\pm$ 1.1\,keVnr.

The PMT-induced noise rejection cut, as detailed in Section~\ref{sec:nai_noise}, utilizing waveform simulation data was implemented. In comparison to the previous selection, this updated criterion demonstrates approximately 20\,\% higher efficiency for events below 1\,keVee with minimal noise contamination. The analysis threshold is set to 5\,NPE corresponding to 0.35\,keVee. 
This enhancement enables the analysis of lower-energy events from smaller scattering angles, which were challenging to extract QFs in the previous measurement.

Assessing the trigger and selection efficiencies is crucial for extracting QFs for low energy nuclear recoils. The efficiency evaluation is reassessed using waveform simulation, and the efficiency is validated with calibration data. During this examination, it is observed that a distinctly lower trigger efficiency, particularly below 2\,keVee ($\approx$30\,NPE), was employed in the previous analysis. This led to a substantial bias toward lower QF values in the low energy nuclear recoil region.

\begin{figure}[!htb]
\centering
\subfigure[]
{
    \includegraphics[width=1.0\columnwidth,keepaspectratio]{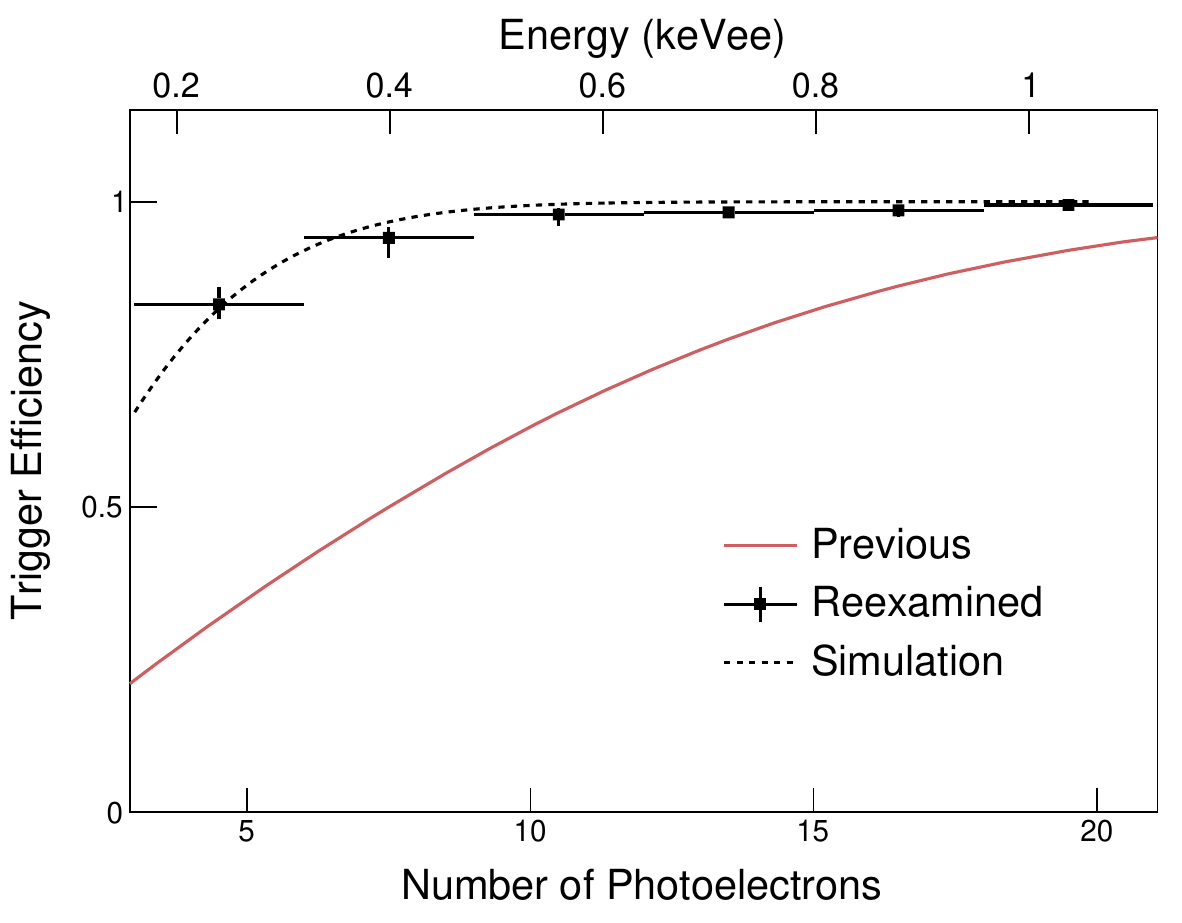}
    \label{fig:rev_eff_trg}
}
\centering
\\
\subfigure[]
{
    \includegraphics[width=1.0\columnwidth,keepaspectratio]{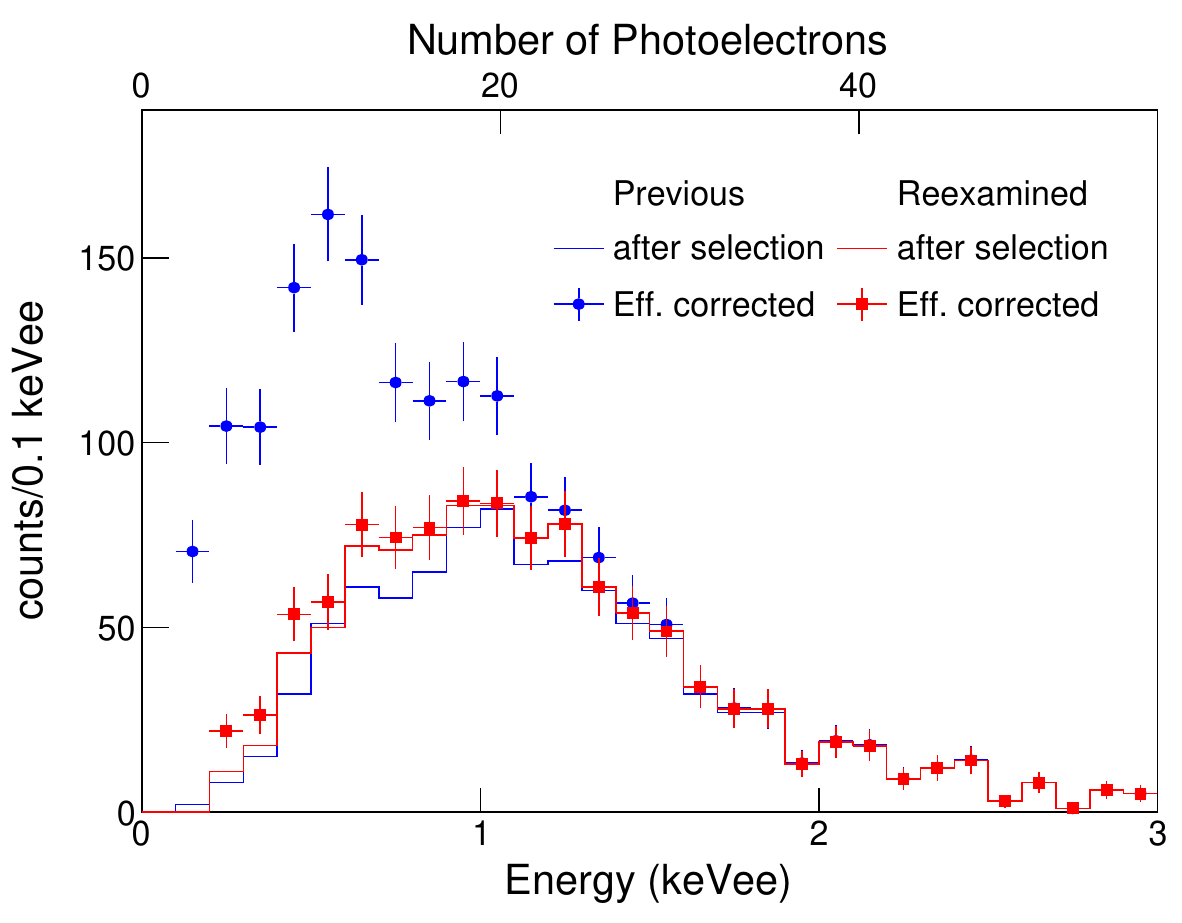}
    \label{fig:rev_eff_e}
}
 \caption{(a) Trigger efficiencies using a $^{22}$Na source with NaI(Tl) and LaBr$_3$ coincident measurement data, comparing the previous analysis (red-solid line) and the re-examined analysis (points), are presented alongside the waveform simulation data (dashed line). The reexamined trigger efficiency aligns well with the results from the waveform simulation data. (b) Energy distributions of the nuclear recoil events tagged by the 16.4$^{\circ}$ neutron tagging detector, corresponding to the nuclear recoil energy of 8.7 $\pm$ 1.3\,keVnr for sodium, are shown. The selected events using the previous analysis (red-solid line) and the re-examined analysis (blue-solid line) are compared with energy spectra after the efficiency correction for the previous analysis (blue-circle points) and the re-examined analysis (red-square points). Due to the skewed trigger efficiency of the previous analysis, the efficiency-corrected spectrum shows an excess of events at low energy, resulting in biased low QF values in the previous analysis.}
\label{fig:rev_eff}
\end{figure}

In the previous measurement, we evaluated the trigger efficiency using an independent setup with NaI(Tl) and LaBr$_{3}$ crystals irradiated by multiple $\gamma$ rays from a $^{22}$Na source~\cite{Joo:2018hom}. The radioactive decay of $^{22}$Na emits three $\gamma$ rays, including one at 1274\,keV and two at 511\,keV. The 511\,keV $\gamma$ from the LaBr${_3}$ crystal was tagged, and low energy events in the NaI(Tl) crystals were identified as coincident Compton scattering events from the other $\gamma$ rays. The measured trigger efficiency was estimated by comparing the total events to the triggered events. 
However, our assumption that all events in the NaI(Tl) crystals coincident with the 511\,keV $\gamma$ in the LaBr${_3}$ crystal were pure scintillation events overlooked the possibility of direct hits on the PMT glasses by $\gamma$ rays, which could generate fast-decaying Cherenkov light. Analysis of the NaI(Tl) and LaBr${_3}$ setup data revealed a sharp increase in the event rate below 1\,keVee, even though 511\,keV coincident $\gamma$ was required in the LaBr$_3$ crystal. Because of the asymmetric behavior of the PMT-induced noise events, the trigger probabilities of these events were significantly lower, resulting in a tendency toward lower trigger efficiency, as illustrated in Fig.~\ref{fig:rev_eff_trg}.



The trigger efficiency was calculated based on the waveform simulation and applied to the QF estimation. For validation of the waveform simulation data, the trigger efficiency was reassessed using the same $^{22}$Na calibration data of the NaI(Tl) and LaBr$_{3}$ setup. PMT-induced noise events were eliminated from the sample, and the increased random event rate from energetic $\gamma$ source irradiation was accounted for by excluding the contribution from random coincidence events. As depicted in Fig.\ref{fig:rev_eff_trg}, the trigger efficiency derived from the waveform simulation aligns well with the reevaluated efficiency. However, the trigger efficiency applied in the previous analysis was markedly lower than that of the re-examined analysis, as well as the waveform simulation data.

Figure~\ref{fig:rev_eff_e} illustrates the energy spectra of nuclear recoil events tagged by the neutron detector at 16.4$^{\circ}$, corresponding to 8.7 $\pm$ 1.3\,keVnr, selected using the previous criteria (blue-solid line) and the re-examined criteria (red-solid line). Additionally, the energy spectra corrected by the evaluated efficiencies for both the previous analysis (blue-circle points) and the re-examined analysis (red-square points) are presented. Due to the biased trigger efficiency in the previous analysis, the QF value was lower, measuring at 9.6 $\pm$ 1.6\%. However, after reexamining the analysis method, the corrected QF is 13.0 $\pm$ 2.2\%.

\begin{figure*}
\centering
\begin{tabular}{cc}
\includegraphics[width=1.0\columnwidth,keepaspectratio]{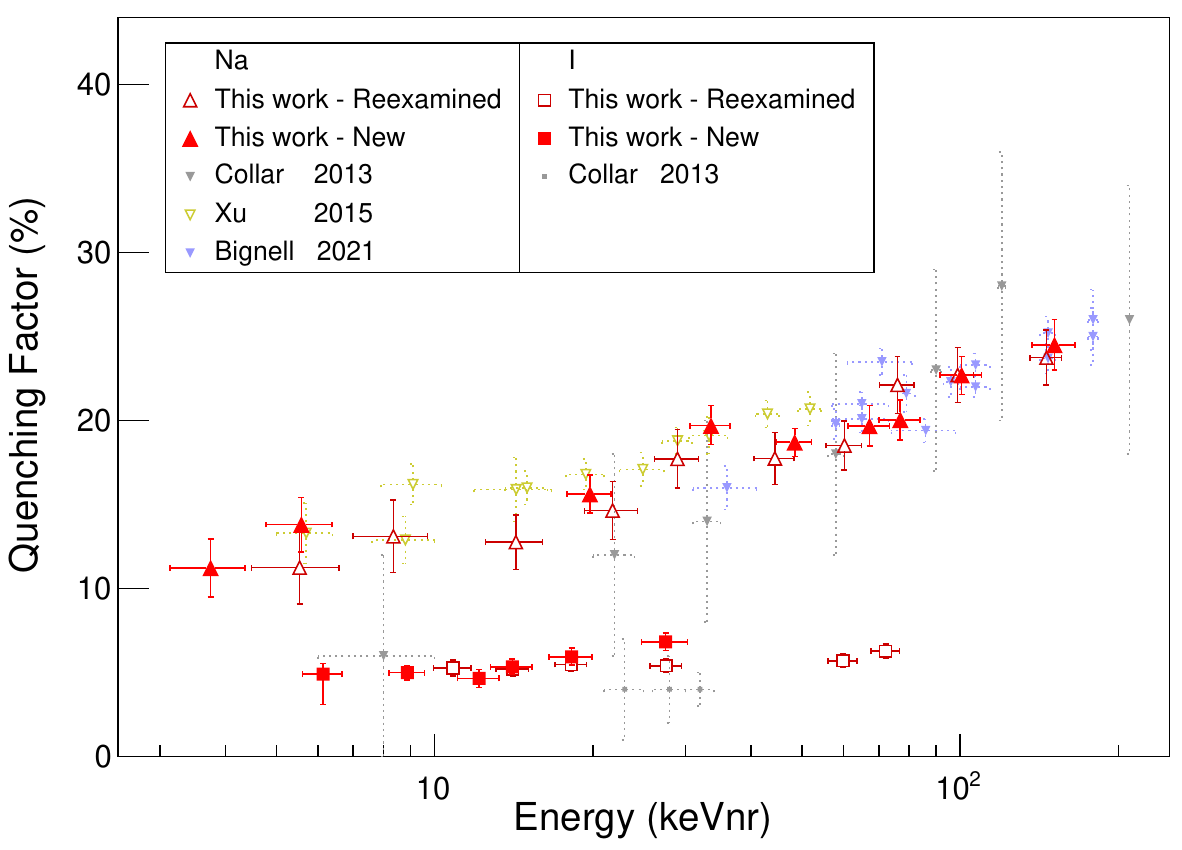}  & 
\includegraphics[width=1.0\columnwidth,keepaspectratio]{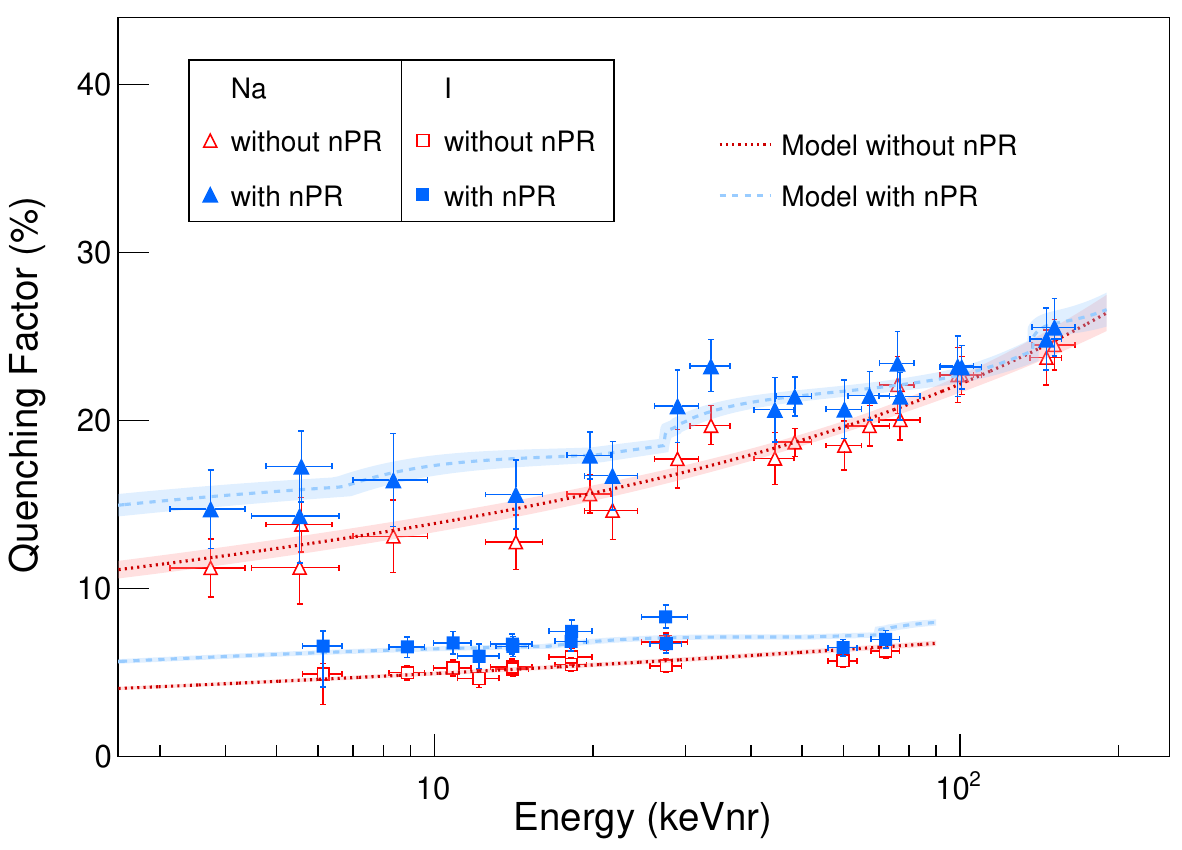} \\
(a) & (b)  \\
\end{tabular}
  \caption{
  (a) Results of the QF measurements for sodium using the high light yield NaI(Tl) crystal (red-filled triangle) and the re-examined analysis of previously reported data (red-open triangle) exhibit consistency with recent measurements from other groups~\cite{sabre1,Bignell:2021bjx}. Similarly, results for iodine from the new measurement (red-filled squares) and the re-examining of previous data (red-open squares) also align with measurements from other study~\cite{Collar:2013gu}. In this new measurement, sodium and iodine QFs were obtained for as low as 3.8\,keVnr and 6.1\,keVnr, respectively.
  (b) The measured QFs (red-open circle for sodium and red-open squares for iodine), calibrated with 59.54\,keV $\gamma$, are modeled using modified Lindhard models (red-dotted lines) with their corresponding uncertainty bands. Taking into account the nonproportionality (nPR) of the NaI(Tl) scintillation process, the measured points are corrected (blue-filled triangles for sodium and squares for iodine). The fitted models are also adjusted accordingly (blue-dashed lines with uncertainty bands).}
\label{fig:QF_result}
\end{figure*}

Table~\ref{tab:reanal} summarizes re-examined QF results in comparison to the previous measurements. The application of advanced event selection has led to the recovery of a few small scattering angles, corresponding to lower recoil energies. Slight increases in QFs with decreased incident neutron energy from new calibration are observed for high scattering angles. Re-examining the trigger efficiency at small scattering angles significantly increases the measured QF values. These updated results align with the findings from the new measurement of the high light yield crystal, as well as with measurements by other groups~\cite{sabre1,Bignell:2021bjx,Collar:2013gu} as shown in Fig.~\ref{fig:QF_result}.

%

\begin{table}[htb]
  \begin{center}
  \caption{Comparison of reexamined nuclear recoil energies ($E_{nr}$) and QFs with previously reported results. The reexamined outcomes consider the neutron beam energy of 2.34 $\pm$ 0.06\,MeV, the application of enhanced event selections, and accurate efficiency evaluation through waveform simulation.}
    \label{tab:reanal}
    \bgroup
    \def\arraystretch{1.3}
      {\footnotesize
      \begin{tabular}{cccccc}
      \hline      \hline
       & Scattering &  \multicolumn{2}{c}{$E_{nr}$ (keVnr)} & \multicolumn{2}{c}{QF (\%)} \\
                    \cline{3-6}
        & angle ($^{\circ}$)	& Prev.	& New &  Prev.	& New \\
      \hline
        Na	& 13.2	&   5.8 $\pm$ 1.0 &   5.5 $\pm$ 1.1	 & 	$-$	 		  & 11.3 $\pm$ 2.2 \\
		& 16.4 	&   8.7 $\pm$ 1.3 &   8.3 $\pm$ 1.4	 &  9.6 $\pm$ 1.6 & 13.0 $\pm$ 2.2 \\
            & 21.3  &  14.8 $\pm$ 1.6 &  14.3 $\pm$ 1.8  & 11.3 $\pm$ 1.2 & 12.8 $\pm$ 1.6 \\
            & 26.6  &  22.7 $\pm$ 2.0 &  21.8 $\pm$ 2.5  & 14.1 $\pm$ 1.3 & 14.6 $\pm$ 1.7 \\
            & 31.0  &  30.1 $\pm$ 2.2 &  29.0 $\pm$ 2.8  & 17.2 $\pm$ 1.3 & 17.7 $\pm$ 1.7 \\
            & 38.2  &  46.1 $\pm$ 2.8 &  44.4 $\pm$ 3.8  & 17.3 $\pm$ 1.1 & 17.7 $\pm$ 1.7 \\
            & 45.0  &  62.6 $\pm$ 3.2 &  60.2 $\pm$ 4.7  & 18.1 $\pm$ 0.9 & 18.5 $\pm$ 1.5 \\
            & 51.3  &  78.9 $\pm$ 3.6 &  76.0 $\pm$ 5.7  & 21.3 $\pm$ 1.0 & 22.1 $\pm$ 1.7 \\
            & 59.0  & 102.7 $\pm$ 4.1 &  98.9 $\pm$ 7.1  & 22.1 $\pm$ 0.9 & 22.7 $\pm$ 1.6 \\
            & 74.7  & 151.6 $\pm$ 5.0 & 145.9 $\pm$ 10.0 & 22.9 $\pm$ 0.7 & 23.7 $\pm$ 1.6 \\
      \hline
       I    & 45.0	 &  11.3 $\pm$ 0.6 & 10.8 $\pm$ 0.9	& 	$-$	 		& 5.3 $\pm$ 0.5 \\
		& 51.3  &  14.6 $\pm$ 0.7 & 14.1 $\pm$ 1.0 &   $-$	 	   & 5.2 $\pm$ 0.4 \\
      	& 59.0  &  18.9 $\pm$ 0.8 & 18.2 $\pm$ 1.3 & 4.3 $\pm$ 0.4 & 5.5 $\pm$ 0.4 \\
            & 74.7  &  28.7 $\pm$ 1.0 & 27.6 $\pm$ 1.9 & 4.7 $\pm$ 0.2 & 5.4 $\pm$ 0.4 \\
            & 126.9 &  62.2 $\pm$ 1.5 & 59.9 $\pm$ 3.8 & 5.6 $\pm$ 0.2 & 5.7 $\pm$ 0.4 \\
            & 159.4 &  74.9 $\pm$ 1.6 & 72.1 $\pm$ 4.5 & 5.9 $\pm$ 0.2 & 6.2 $\pm$ 0.4 \\
        \hline
        \hline
    \end{tabular}
    }
    \egroup
  \end{center}
\end{table}

\section{\label{sec:res}Results and discussions}

QFs for sodium recoils, spanning a broad energy range from 3.8 to 151.6 keVnr, and iodine recoils ranging from 6.1 to 74.9 keVnr, have been successfully obtained, extending this investigation into lower energy regions. Leveraging the high-light-yield crystal, the achievement includes a remarkable measurement of 3.8\,keVnr, representing the lowest nuclear recoil measurement for sodium to date. Through a re-examination of previously reported data, involving updates in neutron energy calibration, event selections, as well as selection and trigger efficiency evaluations, the reexamined QFs are presented alongside the new measurements, as depicted in Fig.~\ref{fig:QF_result}(a). These results are compared with measurements from other research groups~\cite{sabre1,Stiegler:2017kjw,Collar:2013gu}, showing consistent values. The QF values for each recoil angle obtained in this study are detailed in Tables~\ref{tab:QFresult} and ~\ref{tab:reanal}.

The QFs presented in Fig.\ref{fig:QF_result}(a) from various experiments have been linearly calibrated using either a 59.54\,keV $\gamma$ line or a 57.6\,keV $\gamma$ line. However, discrepancies in QF values arise from the utilization of different energy calibration methods involving distinct $\gamma$ energy points, as extensively discussed in Ref.~\cite{Cintas:2024pdu}. This discrepancy is primarily attributed to the nonproportionality of NaI(Tl) crystals~\cite{nPR:489415,nonprop}. To address nonproportionality and ensure an accurate conversion between the measured charge (light output) and deposited energy, it becomes imperative to implement an energy-dependent calibration for both $\gamma$s and electrons.
Our investigation into the nonproportionality phenomenon involved a thorough examination utilizing x rays and $\gamma$ rays sourced from external radioactive materials, as well as internal or cosmogenic contaminants present in the COSINE-100 crystals~\cite{COSINE-100:2024log}. This comprehensive study yielded an energy-dependent calibration curve, enabling the recalculation of QF results. Figure~\ref{fig:QF_result}(b) shows the QF values obtained in this study, accompanied by the recalculated values considering the nonproportionality of the NaI(Tl) crystals for both $\gamma$ and x rays. These data points are modeled using the modified Lindhard model~\cite{Lindhard:1961zz} as described in Ref.~\cite{Ko:2019enb}. The inclusion of nonproportionality in the QF assessment reveals a notable increase in QF values, particularly for the low energy region, underscoring the importance of using appropriate QF values depending on the electron-equivalent energy calibration.

\section{Conclusion}

Measurements of low energy QFs for NaI(Tl) crystals were conducted using a high light yield crystal of 26.0 $\pm$ 0.7\,NPE/keVee, achieved through a novel crystal encapsulation technique. The precise evaluation of QFs for low energy nuclear recoils was facilitated by an in-depth understanding of low energy scintillation events from waveform simulations. These measurements extended to sodium QFs as low as 3.8\,keVnr, marking the first measurement for sodium below 5\,keVnr. Furthermore, a reexamination of previously reported data, with an enhanced comprehension of incident neutron energy and a refined understanding of low energy scintillation events through waveform simulation, was reported. The measurements show general consistency with recent reports from other research groups. These low energy QF measurements for NaI(Tl) crystals hold relevance for future data analyses in the pursuit of dark matter and coherent elastic neutrino-nucleus scattering (CE$\nu$NS) studies with NaI(Tl) crystals, such as the COSINE-100 and NEON experiments.

\begin{acknowledgments}
We thank the IBS Research Solution Center (RSC) for providing high-performance computing resources. 
This work is supported by the Institute for Basic Science (IBS) under project code no. IBS-R016-A1.
\end{acknowledgments}
\bibliographystyle{PRTitle}
\providecommand{\href}[2]{#2}\begingroup\raggedright\endgroup

\end{document}